\documentclass[journal]{IEEEtran}

\usepackage{amssymb}
\newtheorem{theorem}{Theorem}

\newtheorem{remark}{Remark}
\usepackage{cite}
\ifCLASSINFOpdf
  \usepackage[pdftex]{graphicx}
  \graphicspath{{../pdf/}{../jpeg/}}
  \DeclareGraphicsExtensions{.pdf,.jpeg,.png}
\else
  \usepackage[dvips]{graphicx}
  \graphicspath{{../eps/}}
  \DeclareGraphicsExtensions{.eps}
\fi
\usepackage[cmex10]{amsmath}
\usepackage{array}

\usepackage[tight,footnotesize]{subfigure}
\usepackage{microtype}
\usepackage{epstopdf}
\begin{document}
\title{On Index Coding in Noisy Broadcast Channels with Receiver Message Side Information}

\author{\IEEEauthorblockN{Behzad Asadi, Lawrence Ong, and Sarah J.\ Johnson}\thanks{This work is supported by the Australian Research Council under grants FT110100195 and DE120100246.}}

\maketitle
\begin{abstract}
This letter investigates the role of index coding in the capacity of AWGN broadcast channels with receiver message side information. We first show that index coding is unnecessary where there are two receivers; multiplexing coding and superposition coding are sufficient to achieve the capacity region. We next show that, for more than two receivers, multiplexing coding and superposition coding alone can be suboptimal. We give an example where these two coding schemes alone cannot achieve the capacity region, but adding index coding can. This demonstrates that, in contrast to the two-receiver case,  multiplexing coding cannot fulfill the function of index coding where there are three or more receivers.
\end{abstract}


%
\IEEEpeerreviewmaketitle
\vspace{-11pt}
\section{Introduction}
We consider additive white Gaussian noise broadcast channels (AWGN BCs) with receiver message side information, where the receivers know parts of the transmitted messages a priori. Coding schemes commonly used for these channels are \textit{superposition coding}, \textit{multiplexing coding} and \textit{index coding}. In this letter, we show that index coding is redundant where there are only two receivers, but can improve the rate region where there are more than two receivers.
\vspace{-11pt}
\subsection{Background}
Superposition coding is a layered transmission scheme in which the codewords of the layers are linearly superimposed to form the transmitted codeword. Specifically, consider an AWGN channel where the transmitter sends messages $M_1$ and $M_2$. A two-layer transmitted codeword is
\begin{align*}
u^{(n)}\left(M_1\right)+v^{(n)}\left(M_2\right),
\end{align*}
where $u^{(n)}=\left(u_{1}, u_{2},\ldots, u_{n}\right)$ and $v^{(n)}=\left(v_{1}, v_{2},\ldots, v_{n}\right)$ are the codewords of the layers. This scheme was proposed for broadcast channels \cite{BC} and is also used in other channels, e.g., interference channels \cite{InterferenceChannels} and relay channels \cite{RelayChannels}. Superposition coding can achieve the capacity region of degraded broadcast channels \cite{AWGNBCConverse, DMBCConverse}. In broadcast channels with receiver message side information, superposition coding is used in conjunction with multiplexing coding \cite{MultiplexedCoding} and index coding \cite{Index Coding}. 

In multiplexing coding, two or more messages are bijectively mapped to a single message, and a codebook is then constructed for this message. For instance, suppose a transmitter wants to send messages $M_1 \in \{1,2,\dotsc,2^{nR_1}\}$ and $M_2 \in \{1,2,\dotsc, 2^{nR_2}\}$. The single message $M_\text{m}=[M_1,M_2]$ is first formed from $M_1$ and $M_2$, where $[\cdot]$ denotes a bijective map. Then codewords are generated for $M_\text{m}$, i.e., $x^{(n)}\left(m_\text{m}\right)$ where $m_\text{m}\in \{1,2,\ldots,2^{n(R_1+R_2)}\}$. This coding scheme is also called nested coding \cite{NestedCoding} or physical-layer network coding \cite{PHYLayerNetworkCoding}. Multiplexing coding can achieve the capacity region of broadcast channels where the receivers know some of the messages demanded by the other receivers, and want to decode all messages \cite{BCwithSI2UsersOechtering}. This scheme can also achieve the capacity region of a more general scenario where a noisy version of messages is available at the receivers, who also need to decode all messages \cite{SWoverBC}.

The transmitter of the broadcast channel can utilize the structure of the side information available at the receivers to accomplish compression by XORing its messages. This is performed such that the receivers can recover their requested messages using XORed messages and their own side information. As an example, consider a broadcast channel where $M_1$ and $M_2$ are the two requested messages by two receivers, and each receiver knows the requested message of the other receiver. Then the transmitter only needs to transmit $M_\text{x}=M_1\oplus M_2$, which is the bitwise XOR of $M_1$ and $M_2$ with zero padding for messages of unequal length. Modulo-addition can also be used instead of the XOR operation \cite{BCwithSI2UsersKramer}. This coding scheme is called index coding. It is also called network coding \cite{NetworkCoding}, as any index coding problem can be formulated as a network coding problem\cite{NetworkIndexCoding}.

Separate index and channel coding is a suboptimal scheme in broadcast channels with receiver message side information. The separation, where the receiver side information is not considered during the channel decoding of the XOR of the two messages, leads to a strictly smaller achievable rate region in two-receiver broadcast channels \cite{BCwithSI2UsersOechtering, PHYLayerNetworkCoding}.  Separate index and channel coding has been shown to achieve within a constant gap of the capacity region of three-receiver AWGN BCs with only private messages \cite{BCwithSI3UsersPrivateMessage}.

Using a combination of index coding, multiplexing coding and superposition coding, Wu \cite{BCwithSI2UsersGeneral} characterized the capacity region of two-receiver AWGN BCs for all possible message and side information configurations. In this setting, the first receiver requests $\{M_1,M_3,M_5\}$ and knows $M_4$, and the second receiver requests $\{M_2,M_3,M_4\}$ and knows $M_5$. The transmitted codeword of the capacity-achieving transmission scheme is
\vspace{-5pt}
\begin{align}\label{WuScheme}
u^{(n)}\left(M_1\right)+v^{(n)}\left(M_\text{mx}\right),
\end{align}
where $M_\text{mx}=[M_2,M_3,M_4\oplus M_5]$.

The combination of these coding schemes can also achieve the capacity region of some classes of three-receiver less-noisy and more-capable broadcast channels where (i) only two receivers possess side information, and (ii) the only message requested by the third receiver is also requested by the other two receivers (i.e, a common message)\cite{BCwithSI3UsersCommonMessage}.
\subsection{Contributions}
In this work, we first show that, in two-receiver AWGN BCs with receiver message side information, index coding need not necessarily be applied prior to channel coding if multiplexing coding is used. To this end, we show that index coding is a redundant coding scheme in (\ref{WuScheme}), and multiplexing coding and superposition coding are sufficient to achieve the capacity region. We then derive the capacity region of a three-receiver AWGN BC. Prior to this letter, the best known achievable region for this channel was within a constant gap of the capacity region~\cite{BCwithSI3UsersPrivateMessage}. In this channel, index coding proves to be useful in order to achieve the capacity region; superposition coding and multiplexing coding cannot achieve the capacity region of this channel without index coding. Our result indicates that index coding cannot be made redundant by multiplexing coding in broadcast channels with receiver message side information where there are more than two receivers.
\vspace{-5pt}
\section{AWGN BC with Side Information}\label{AWGN BC with SI}
\begin{figure}[t]
\centering
\includegraphics[width=0.42\textwidth]{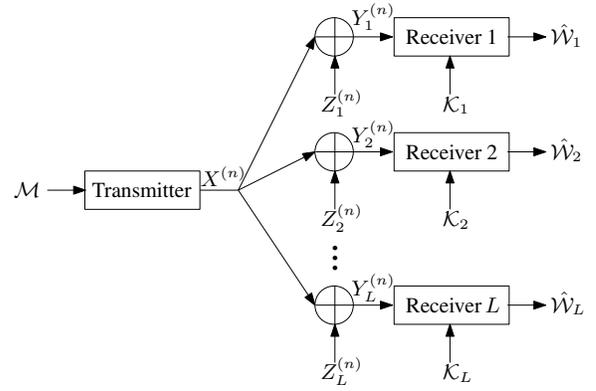}
\vspace{-5pt}
\caption{The AWGN broadcast channel with receiver message side information, where $\mathcal{W}_i \subseteq \mathcal{M}$ is the set of messages demanded by receiver $i$, and  $\mathcal{K}_i \subset \mathcal{M}$ is the set of messages known to receiver $i$ a priori.} 
\vspace{-5pt}
\label{AWGNBCModelFig}
\end{figure}

In an ${L}$-receiver AWGN BC with receiver message side information, as depicted in Fig. \ref{AWGNBCModelFig}, the signals received by receiver $i$, $Y_{i}^{(n)}\;i=1,2,\ldots,L$, is the sum of the transmitted codeword, $X^{(n)}$, and an i.i.d. noise sequence, $Z_i^{(n)} \;i=1,2,\ldots,L$, with normal distribution, $Z_i\sim \mathcal{N}\left(0, N_i\right)$. The transmitted codeword has a power constraint of $\sum_{l=1}^{n}E\left(X_l^2\right)\leq nP$ and is a function of source messages, $\mathcal{M}=\{M_1,M_2,\ldots,M_K\}$. The messages $\{M_j\}_{j=1}^K$ are independent, and each message, $M_j$, is intended for one or more receivers at rate $R_j$. This channel is stochastically degraded, and without loss of generality, we can assume that receiver $1$ is the strongest and receiver $L$ is the weakest in the sense that $N_1\leq N_{2} \leq \cdots \leq N_L$. 

To model the request and the side information of each receiver, we define two sets corresponding to each receiver; the \textit{wants} set, $\mathcal{W}_i$, is the set of messages demanded by receiver $i$ and the \textit{knows} set, $\mathcal{K}_i$, is the set of messages that are known to receiver $i$.
\vspace{-5pt}
\section{Where Index Coding is Not Required}\label{Two-user AWGN BC with SI}
In this section, we consider the general message setting in two-receiver AWGN BCs with receiver message side information, where the \textit{knows} and \textit{wants} sets of the receivers are given by 
\begin{equation}\label{tworeceiverchannel}
\begin{split} 
&\text{Receiver 1: } \mathcal{W}_1=\{M_1,M_3,M_5\}, \mathcal{K}_1=\{M_4\},\\
&\text{Receiver 2: } \mathcal{W}_2=\{M_2,M_3,M_4\}, \mathcal{K}_2=\{M_5\}.
\end{split}
\end{equation}

The capacity region of this channel has been derived by Wu~\cite{BCwithSI2UsersGeneral}. It is achievable using a combination of index coding, multiplexing coding, and superposition coding, as in \eqref{WuScheme}.

For the special case where $M_1 = M_2 = M_3 = 0$, Oechtering et al.~\cite{BCwithSI2UsersOechtering} have shown that multiplexing coding alone can achieve the capacity region. In spirit of Oechtering et al., we now show that index coding is also not necessary for the general case \eqref{tworeceiverchannel}. 
\begin{theorem}
Multiplexing coding and superposition coding are sufficient to achieve the capacity region of two-receiver AWGN BCs with receiver message side information.
\end{theorem}

\begin{IEEEproof}
We can use only multiplexing coding and superposition coding to achieve the capacity region of the two-receiver AWGN BC with the general message setting given in \eqref{tworeceiverchannel}, i.e.,
\begin{align}\label{WuSchemewithoutindexcoding}
x^{(n)}=u^{(n)}\left(M_1\right)+v^{(n)}\left(M_\text{m}\right),
\end{align}
where $M_\text{m}=[M_2,M_3,M_4,M_5]$. The only difference between \eqref{WuScheme} and \eqref{WuSchemewithoutindexcoding} is that $M_4\oplus M_5$ in $M_\text{mx}$ is replaced with  $[M_4,M_5]$. This scheme achieves the same rate region as \eqref{WuScheme}, i.e., the capacity region. This is because from the standpoint of each receiver, the amount of uncertainty to be resolved in $M_4\oplus M_5$  is the same as that in  $[M_4,M_5]$. 
This uncertainty to be resolved in $M_4\oplus M_5$ (or $[M_4,M_5]$)  is $M_5$ by receiver 1, and $M_4$ by receiver 2.
\end{IEEEproof}

This result indicates that multiplexing coding can fulfill the function of index coding in two-receiver AWGN BCs with receiver message side information.
\begin{remark}
For two-receiver \textit{discrete memoryless} broadcast channels with receiver message side information, we conjecture that index coding is not necessary. Using the same reasoning as above, we can always replace it with multiplexing coding without affecting the achievable rate region.
\end{remark}
\section{Where Index Coding Improves the Rate Region}\label{three-receiver AWGN BC with SI}
In this section, we demonstrate that multiplexing coding cannot fulfill the role of index coding in AWGN broadcast channels with receiver message side information where there are more than two receivers. To this end, we establish the capacity region of a three-receiver AWGN BC where the \textit{wants} and \textit{knows} sets of the receivers are
\begin{equation}\label{threeuserchannel}
\begin{split} 
&\text{Receiver 1: } \mathcal{W}_1=\{M_1\}, \mathcal{K}_1=\varnothing,\\
&\text{Receiver 2: } \mathcal{W}_2=\{M_2\}, \mathcal{K}_2=\{M_3\},\\
&\text{Receiver 3: } \mathcal{W}_3=\{M_3\}, \mathcal{K}_3=\{M_2\},
\end{split}
\end{equation}
and show that multiplexing coding and superposition coding cannot achieve the capacity region of this channel without index coding. 
\vspace{-5pt}
\subsection{Using Index Coding Prior to Multiplexing Coding and Superposition Coding}
In this subsection, we establish the capacity region of the broadcast channel of interest, stated as Theorem \ref{maintheorem}.
\begin{theorem}\label{maintheorem}
The capacity region of the three-receiver AWGN BC with the configuration given in (\ref{threeuserchannel}) is the closure of the set of all rate triples $(R_1,R_2,R_3)$, each satisfying
\begin{subequations}
\begin{align}
&R_1 < C\left(\frac{\alpha P}{N_1}\right),\label{CapacityR1}\\
&R_2 < C\left(\frac{(1-\alpha)P}{\alpha P+N_2}\right),\label{CapacityR2}\\
&R_3 < C\left(\frac{(1-\alpha)P}{\alpha P+N_3}\right),\label{CapacityR3}
\end{align}
\end{subequations}
for some $0\leq\alpha\leq1$, where $C(x)=\frac{1}{2}\log_2(1+x)$. 
\end{theorem}
Here, we prove the achievability part of Theorem \ref{maintheorem}; the proof of the converse is presented in the appendix.
\begin{IEEEproof}(Achievability)
Index coding and superposition coding are employed to construct the transmission scheme that achieves the capacity region. The codebook of this scheme contains two subcodebooks. The first subcodebook includes $2^{nR_1}$ i.i.d. codewords, $u^{(n)}\left(m_1\right)$ where $m_1\in \{1,2,\ldots,2^{nR_1}\}$, and $U\sim \mathcal{N}\left(0, \alpha P\right)$ for an $0\leq \alpha \leq 1$. The second subcodebook includes $2^{n\max \{R_2,R_3\}}$ i.i.d. codewords, $v^{(n)}\left(m_\text{x}\right)$ where $m_\text{x}=m_2\oplus m_3$, $m_\text{x}\in \{1,2,\ldots,2^{n\max\{R_2,R_3\}}\}$, $V\sim \mathcal{N}\left(0, (1-\alpha) P\right)$, and $V$ is independent of $U$. Using superposition coding, the transmitted codeword over the broadcast channel is given by
\begin{equation}\label{indexcodingscheme}
x^{(n)}=u^{(n)}\left(M_1\right)+v^{(n)}\left(M_\text{x}\right).
\end{equation}
The achievability of the region in (\ref{CapacityR1})-(\ref{CapacityR3}) using the transmission scheme in (\ref{indexcodingscheme}) can be verified by considering two points during the decoding. First, receivers 2 and 3 consider $u^{(n)}$ as noise. Since receiver 2 knows $M_3$ a priori, and receiver 3 knows $M_2$ a priori, we obtain (\ref{CapacityR2}) and (\ref{CapacityR3}) as the requirements for achievability. Second, receiver 1 decodes $m_\text{x}$ while treating $u^{(n)}$ as noise. This requires
\begin{align}\label{redundantcondition}
\max\{R_2,R_3\} < C\left(\frac{(1-\alpha)P}{\alpha P+N_1}\right),
\end{align}
for achievability. However, considering the inequalities in (\ref{CapacityR2}) and (\ref{CapacityR3}), this condition is redundant and can be dropped. Receiver 1 then removes $v^{(n)}$ from its received signal and decodes $m_1$, which yields (\ref{CapacityR1}) in the achievable region.
\end{IEEEproof}
\vspace{-4pt}
\subsection{Not Using Index Coding Prior to Multiplexing Coding and Superposition Coding}
In this subsection, we characterize the achievable rate region for the broadcast channel of interest when $M_\text{x}=M_2\oplus M_3$ in \eqref{indexcodingscheme} is replaced with $M_\text{m}=[M_2,M_3]$. This employs the same XOR-multiplexing substitution shown to be optimal in the two-receiver case. This means that the messages are directly fed to multiplexing coding and superposition coding. The codebook of this transmission scheme also contains two subcodebooks in which only the second subcodebook is different from the scheme using index coding. The second subcodebook of this scheme includes $2^{n(R_2+R_3)}$ i.i.d. codewords, $v^{(n)}\left(m_\text{m}\right)$ where $m_\text{m}=[m_2,m_3]$, $m_\text{m}\in \{1,2,\ldots,2^{n(R_2+R_3)}\}$, $V\sim \mathcal{N}\left(0, (1-\alpha) P\right)$ for an $0\leq \alpha \leq 1$, and $V$ is independent of $U$. The transmitted codeword of this scheme using superposition coding is given by
\begin{equation}\label{group2scheme}
x^{(n)}=u^{(n)}\left(M_1\right)+v^{(n)}\left(M_\text{m}\right).
\end{equation}

The requirements for achievability concerning the decoding at receivers 2 and 3 are the same as (\ref{CapacityR2}) and (\ref{CapacityR3}). This is because the unknown information rates of both $M_\text{x}$ and $M_\text{m}$ are the same from the standpoint of each of these receivers. So, as far as these two receivers are concerned, multiplexing coding gives the same result as index coding. 

However, decoding $v^{(n)}$ at receiver 1 while treating $u^{(n)}$ as noise requires
\begin{align}\label{extracondition}
R_2+R_3 < C\left(\frac{(1-\alpha)P}{\alpha P+N_1}\right),
\end{align}
for achievability which is not a redundant condition considering (\ref{CapacityR2}) and (\ref{CapacityR3}). The difference between \eqref{redundantcondition} and \eqref{extracondition} is because receiver 1 needs to decode the correct $v^{(n)}$ over the set of $2^{n\max\{R_2,R_3\}}$ candidates for the former, but over the set of $2^{n(R_2+R_3)}$ candidates for the latter. After decoding $v^{(n)}$, receiver 1 decodes $m_1$ which requires (\ref{CapacityR1}) for achievability. When the messages are directly fed to  multiplexing coding and superposition coding, an extra condition, given in (\ref{extracondition}), is required for achievability and, as a result, this scheme cannot achieve the capacity region. Even if receiver 1 does not decode $v^{(n)}$ and treats it as noise, it can only decode $u^{(n)}$ at rates up to $R_1 < C\left(\frac{\alpha P}{(1-\alpha)P+N_1}\right)$, which is strictly smaller than \eqref{CapacityR1}. Receiver 1 can also use simultaneous decoding {\cite[p. 88]{NITBook}} to decode $m_1$ which requires \eqref{CapacityR1} and $R_1+R_2+R_3 < C\left(\frac{P}{N_1}\right)$ for achievability; in this case, the extra condition on the sum rate prevents this scheme from achieving the capacity region.

Note that alternative message combinations for multiplexing coding and superposition coding are also possible. The proof of their suboptimality is straightforward but tedious and repetitive.
\section{Conclusion}\label{Conclusion}
In this work, we first showed that multiplexing coding can fulfill the function of index coding in two-receiver AWGN BCs with receiver message side information. We next established the capacity region of a three-receiver AWGN BC, where superposition coding and multiplexing coding alone cannot achieve the capacity region unless index coding is also used. This shows that index coding cannot be discharged by multiplexing coding in broadcast channels with receiver message side information where there are more than two receivers.
\vspace{-5pt}
\section*{Appendix}
Based on the proofs for the AWGN BC without side information \cite{AWGNBCConverse, NITBook}, we prove the converse part of Theorem \ref{maintheorem} using Fano's inequality and the entropy power inequality (EPI). 
We also use the fact that the capacity region of a stochastically degraded broadcast channel without feedback is the same as its equivalent physically degraded broadcast channel \cite[p. 444]{NITBook} where the channel input and outputs form a Markov chain, $X\rightarrow Y_{1}\rightarrow Y_{2} \rightarrow \cdots \rightarrow Y_{L}$, i.e.,
\begin{align}
&Y_1=X+Z_1,\nonumber\\
&Y_i=Y_{i-1}+\tilde{Z}_i \;\;\;i=2,3,\ldots,L\label{noisetilde},
\end{align}
 where $\tilde{Z}_i \sim \mathcal{N}\left(0, N_i-N_{i-1}\right)$ for $i=2,3,\ldots,L$.
\begin{IEEEproof}(Converse) Based on Fano's inequality, we have
\begin{align}
H\left(M_1 \mid Y_1^{(n)}\right)\leq n\epsilon_n,\label{fano1}\\
H\left(M_2 \mid Y_2^{(n)},M_3\right)\leq n\epsilon'_n,\label{fano2}\\
H\left(M_3 \mid Y_3^{(n)},M_2\right)\leq n\epsilon''_n,\label{fano3}
\end{align}
where $\epsilon_n$, $\epsilon'_n$ and $\epsilon''_n$ tend to zero as $n\rightarrow \infty$. For the sake of simplicity we use $\epsilon_n$ instead of $\epsilon'_n$ and $\epsilon''_n$ as well in the rest. The rate $R_2$ is upper bounded as 
\begin{align}\label{proof11}
&nR_2=H(M_2)\nonumber\\
&=H(M_2 \mid Y_2^{(n)},M_3)+I(M_2;Y_2^{(n)},M_3) \nonumber\\
&\overset{(a)}{=}H(M_2 \mid Y_2^{(n)},M_3)+I(M_2;Y_2^{(n)} \mid M_3) \nonumber\\
&=H(M_2 \mid Y_2^{(n)},M_3)+h(Y_2^{(n)}\mid M_3)-h(Y_2^{(n)}\mid M_2,M_3) \nonumber\\
&\overset{(b)}{\leq}n\epsilon_n+h(Y_2^{(n)}\mid M_3)-h(Y_2^{(n)}\mid M_2,M_3) \nonumber\\
&\overset{(c)}{\leq}n\epsilon_n+\frac{n}{2}\log2\pi e(P+N_2)-h(Y_2^{(n)}\mid M_2,M_3) \nonumber\\
&\overset{(d)}{=}n\epsilon_n+\frac{n}{2}\log2\pi e(P+N_2)-\frac{n}{2}\log2\pi e(\alpha P+N_2),
\end{align}
where $(a)$ follows from the independence of $M_2$ and $M_3$, $(b)$ from (\ref{fano2}), and $(c)$ from $h(Y_2^{(n)}\mid M_3)\leq h(Y_2^{(n)})\leq \frac{n}{2}\log2\pi e(P+N_2)$. In \eqref{proof11}, $(d)$ is from the fact that
\begin{align*}
&\frac{n}{2}\log2\pi e N_2=h(Z_2^{(n)})\\
&=h(Y_2^{(n)}\mid X^{(n)})\overset{(e)}{=}h(Y_2^{(n)}\mid M_2,M_3,X^{(n)})\\
&\leq h(Y_2^{(n)}\mid M_2,M_3)\leq h(Y_2^{(n)}) \leq \frac{n}{2}\log2\pi e(P+N_2),
\end{align*}
where (e) is because $\left(M_2,M_3\right) \rightarrow X^{(n)} \rightarrow Y_2^{(n)}$ form a Markov chain; then since $\frac{n}{2}\log2\pi e N_2\leq h(Y_2^{(n)}\mid M_2,M_3)\leq \frac{n}{2}\log2\pi e(P+N_2)$, there must exist an $0\leq \alpha \leq1$ such that
\begin{align}\label{alphadefinition}
h(Y_2^{(n)}\mid M_2,M_3)=\frac{n}{2}\log2\pi e(\alpha P+N_2).
\end{align}
In this channel, $R_3$ is also upper bounded as
\begin{align}\label{proof12}
&nR_3=H(M_3)\nonumber\\
&=H(M_3 \mid Y_3^{(n)},M_2)+I(M_3;Y_3^{(n)}\mid M_2) \nonumber\\
&\leq n\epsilon_n+h(Y_3^{(n)}\mid M_2)-h(Y_3^{(n)}\mid M_2,M_3) \nonumber\\
&\leq n\epsilon_n+\frac{n}{2}\log2\pi e(P+N_3)-h(Y_3^{(n)}\mid M_2,M_3) \nonumber\\
&\overset{(f)}{\leq}n\epsilon_n+\frac{n}{2}\log2\pi e(P+N_3)-\frac{n}{2}\log2\pi e(\alpha P+N_3),
\end{align}
where $(f)$ is the result of substituting from (\ref{alphadefinition}) and 
\begin{align*}
h(\tilde{Z}_3^{(n)}\mid M_2,M_3)=h(\tilde{Z}_3^{(n)})=\frac{n}{2} \log 2 \pi e\left(N_3-N_2\right),
\end{align*}
into the conditional EPI {\cite[p. 22]{NITBook}} for $Y_3^{(n)}=Y_2^{(n)}+\tilde{Z}_3^{(n)}$ where we have
\begin{align*}
2^{\frac{2}{n}h\left(Y_3^{(n)}\mid M_2,M_3 \right)} \geq 2^{\frac{2}{n}h\left(Y_2^{(n)}\mid M_2,M_3 \right)}+2^{\frac{2}{n}h\left(\tilde{Z}_3^{(n)}\mid M_2,M_3 \right)}.
\end{align*}
Finally, for $R_1$, we have
\vspace{-5pt}
\begin{align}\label{proof13}
&nR_1=H(M_1)\nonumber\\
&=H(M_1 \mid Y_1^{(n)},M_2,M_3)+I(M_1;Y_1^{(n)} \mid M_2,M_3) \nonumber\\
&\leq n\epsilon_n+h(Y_1^{(n)}\mid M_2,M_3)-h(Y_1^{(n)}\mid M_1,M_2,M_3) \nonumber\\
&\overset{(g)}{\leq} n\epsilon_n+\frac{n}{2}\log2\pi e(\alpha P+N_1)-h(Y_1^{(n)}\mid M_1,M_2,M_3) \nonumber\\
&\overset{(l)}{=}n\epsilon_n+\frac{n}{2}\log2\pi e(\alpha P+N_1)-\frac{n}{2}\log2\pi e N_1,
\end{align}
where $(g)$ is the result of substituting from \eqref{alphadefinition} and 
\begin{align*}
h(\tilde{Z}_2^{(n)}\mid M_2,M_3)=h(\tilde{Z}_2^{(n)})=\frac{n}{2} \log 2 \pi e\left(N_2-N_1\right),
\end{align*}
into the conditional EPI for $Y_2^{(n)}=Y_1^{(n)}+\tilde{Z}_2^{(n)}$. 
In (\ref{proof13}), $(l)$ is due to
\vspace{-5pt}
\begin{multline*}
h(Y_1^{(n)}\mid M_1,M_2,M_3)\\
=h(Y_1^{(n)}\mid X^{(n)})=h(Z_1^{(n)})=\frac{n}{2}\log2\pi e N_1.
\end{multline*}

From (\ref{proof11}), (\ref{proof12}), (\ref{proof13}) and since $\epsilon_n$ goes to zero as $n \rightarrow \infty$, the proof of the converse for Theorem \ref{maintheorem} is complete.
\end{IEEEproof}
%
\vspace{-5pt}
\bibliographystyle{IEEEtran}

\end{document}